# LIBRA: IS IT REALLY ABOUT MONEY?

**Valerie Khan**
Vice President
Digital Equity Association
valerie.khan@d-eq.org

**Geoffrey Goodell**
Centre for Blockchain Technologies
University College London
g.goodell@ucl.ac.uk

**Abstract**

The announcement by Facebook that Libra will "deliver on the promise of 'the internet of money'" has drawn the attention of the financial world. Regulators, institutions, and users of financial products have all been prompted to react and, so far, no one managed to convince the association behind Libra to apply the brakes or to convince regulators to stop the project altogether. In this article, we propose that Libra might be best seen not as a financial newcomer, but as a critical enabler for Facebook to acquire a new source of personal data. By working with financial regulators seeking to address concerns with money laundering and terrorism, Facebook can position itself for privileged access to high-assurance digital identity information. For this reason, Libra merits the attention of not only financial regulators, but also the state actors that are concerned with reputational risks, the rule of law, public safety, and national defence.

## Introduction

These days we are often too impatient to read a book or an article from beginning to end. But in today's short attention-span culture, it might be ever more important to ensure you don't miss out on the final message – or maybe something that was intentionally buried in a document to be hidden. In movies, this would be called a prolepsis: a scene that temporarily jumps the narrative forward in time.

In the case of Libra, the prolepsis can be found in Section 5: "An additional goal of the association is to develop and promote an open identity standard. We believe that decentralized and portable digital identity is a prerequisite to financial inclusion and competition." [1]

That is not to say that launching an association of members that aims to create "a reliable digital currency and infrastructure that together can deliver on the promise of 'the internet of money'" is not a massive statement. It is! But it also keeps us busy trying to think about the reaction of regulators and banks: how will China [2] respond and how will central banks deal with the accumulation of assets to guarantee a stable value for their coin-to-be? Will it still be possible to tax a transaction? These are all important questions. But what if this is a decoy?

What if there is something more? What if Libra is actually aiming to own the solution to the even bigger and older problem of digital identity? As the classic "New Yorker" cartoon put it, "on the internet, nobody knows you're a dog" [3].

Analysing this question will unfold the massive potential of this space, and its specific interest for an advertising company like Facebook. Allowing Facebook to become a crucial player in digital identity for the financial sector will enable it to tighten the knot on the 'transparent citizen' [4] by accessing a strong bastion of meaningful data. It will also allow everyone else to purchase the means to manipulate Facebook users, perhaps in pursuit of their respective advertising ideas – some harmless, some of corrupting influence. By dressing this up as a financial inclusion project, Facebook manages to draw financial services regulators to the table. Yet, this should also call everyone who is looking at reputational risks, the rule of law, public safety, and national defence. Otherwise, states and societies might just be designing their Maginot Line.

## The identity dilemma

With the Internet becoming increasingly engrained in our day-to-day lives, issues relating to our online identities have become more important, and the potential impact on our lives has become more significant. We as individuals interact with a number of online sites and services which require an account to access, typically with a username and a password [5]. This could be something as simple as news sites that restrict the number of articles you can read anonymously in a certain time period, or it could be something as personal as your financial or healthcare services. Depending on the required degree of assurance, the type of information collected to establish such accounts might include verifiable email addresses or phone numbers, which could be used to establish profiles, as well as signed credentials by which certified authorities can confer trust. In some instances, this might also contain biometric information. In an analogue world, we would typically provide this information directly to human representatives of the service provider, who would generally not copy down the details: Trust was created through the personal contact. However, in the online world we increasingly call on third-party providers, like Facebook, to help us open the door and interact with our service providers. This process is referred to as authentication, meaning that someone (either the site themselves or a third party) is making sure that the user accessing is indeed who he or she claims to be.

The question of trusting this authentication process has become one of the most prevailing questions of our digital times. The dilemma begins on a very personal level. Through different channels, online and offline, private and public companies build up numerous identifiers for each single user. In time, we are broken down into silos: we become different persons in each channel, represented by an email ID, a device ID, a customer login, or a 1st party cookie. Yet, for a reliable authentication, the service provider will need (or want) to re-establish sufficient amount of information about the user.

With the rise of Big Data analytics, data brokers can acquire information through a variety of channels. Facebook ranks among the most preeminent data brokers. The company has multiple data points about its users, potentially across various apps. According to the Washington Post [6], Facebook aims to collect at least 98 data points for the purpose of ensuring that targeted ads are "useful and relevant". This includes demographics like your age and ethnicity, your on-site activities such as the pages you like and the ads you click, and your device and location settings such as the brand of phone you use and your type of Internet connection. But it also includes Facebook's web-tracking efforts and its collaborations with major data brokers like Experian, Acxiom and Epsilon. In some cases, even offline data are being added to complete the picture. Data collection, storage, and processing are ubiquitous in the networked ecosystem.

What Facebook is doing is to re-identify [7] the fragmented online information and linking the data back to one unitary person that can be targeted, profiled and surveilled. This is typically referred to as record linking or entity resolution [8]: the task of finding records in a data set that refer to the same entity (in this case identity) across different data sources, often through a common identifier like a phone number, a national ID, or a date of birth. Such linking forces individuals into a perpetual tracking mechanism that will create the 'transparent citizen'. Daniel J. Solove, a law professor at George Washington University, argues: "In the individual case, it might be innocuous, say the fact that I like Coke and not Pepsi [but] when you put together pieces of information, it becomes different. When you have a record of everything you ever bought over the years, you can make inferences about your health, financial situation and interests." [9]

This 'transparent citizen' enables Facebook to more accurately determine the truth of the information provided – at least this seems to be the growing perception. Tacitly, this way we assume that everyone might be trying to commit fraud. Concomitantly, it suggests that the more transparent we are, the more trustworthy we are. The conclusion that the exigency for authentication might justify the means of data collection relies upon the assumption that the human costs of holding substantially all forms of interaction to the standards of high assurance can be safely ignored. This assumption puts forward the question of power over declaring us rightful or not. It is self-evident that the insistence upon universal trustworthiness forces everyone to disclose everything, which in turn becomes the opposite of the freedom of opinion. This was established in 1995, when the United States Supreme Court in *McIntyre v. Ohio Elections Commission* determined: "Anonymity is a shield from the tyranny of the majority." [10]

## A matter of power

The ability to force individuals to justify their trustworthiness with high assurance, to whichever third parties may ask, is a position of true power indeed. It is worthwhile to ask whether an implicit goal of the data brokers is to force individuals into a position of global consistency in which they must each present the same face to everyone. Mark Zuckerberg explained his view during an interview in 2009: "You have one identity [...] The days of you having a different image for your work friends or co-workers and for the other people you know are probably coming to an end pretty quickly ... Having two identities for yourself is an example of a lack of integrity." [11] In a later quote, Mark Zuckerberg explains further with the forgone conclusion: "[…] the level of transparency the world has now won't support having two identities for a person." [11] On this basis, letting a concentrated group of revenue-seeking organisations define the de facto standards and policies, would be like allowing a single family with the requisite printing facilities to manage the issuance of passports. Standards can be used as a vector for directing the manner by which groups interact. Legitimate standards organisations endeavour to produce neutral standards since the standards they produce ensure the requirements of interoperability. If they are not neutral, how long would it take until decisions are being taken to exclude individuals or groups, until information is being captured that can harm people and lead to repercussions, or until the right to have a passport would be too expensive for some citizens?  And if they define the standards, who will control them? And how will competitors be able to enter the market? During the US House of Representatives Financial Services Committee on the 17th of July [12], Facebook's crypto chief David Marcus, was asked by Rep. Sean Duffy, R-WI, if people who have been banned from Facebook's social network like Milo Yiannopoulos and Louis Farrakhan [13] would be allowed to use Libra. The answer was: "We haven't written the policy yet," indicating that the rules of entry will be written by Facebook. But what business does Facebook have in setting public policy anyway? Whence does its legitimacy arise? Concerns about power-misuse of a public good like money, as highlighted by Alexandria Ocasio-Cortez, are legitimate [14].

Before [15] and during the US House Committee hearing [12], the Committee called on Facebook to agree to a moratorium to allow time for "examining Facebook's proposed cryptocurrency and its impact on consumers, investors, and the American financial system". Facebook's answer was very accommodating as noted by CNBC: "We understand that big ideas take time, that policymakers and others are raising important questions, and that we can't do this alone," the Facebook executive wrote [16]. "We want, and need, governments, central banks, regulators, non-profits, and other stakeholders at the table and value all of the feedback we have received." This apparent magnanimity indicates that Facebook is setting the terms of engagement. But equally, the regulator cannot stop Facebook from offering their wallet, called Calibra, that allows for payments within a closed-loop system (as is currently planned for all, but only, members of the association) [17]. Although financial service regulators will watch closely to ensure that they are not enabling a payment mechanism that can serve illicit payments and require a control mechanism for which payment providers have to collect information about the buyer, the seller and the transaction (according to FATF Recommendations [18]), Facebook has little to fear. No one expects that a payment authorised by Facebook's infrastructure would be private, and this is precisely what makes it so dangerous. The risk-aversion surrounding private payment mechanisms plays right into the hands of Facebook executives, who can appease financial regulators seeking an innovative solution to rein in money laundering and terrorist financing by building a high-assurance surveillance system without public funds.

## Tell me who you are, and I tell you what you want

Regulators will demand Facebook to keep information from Calibra separate to the rest of the Facebook database, but we need to be clear on what we mean by information. There will be data concerning transactions and accounts, and there will be data around the unitary identity of the client owning the wallet. Whilst most will demand the transaction data to be hidden, it will become a natural part of the identity data: identity is really just the linkage of transactions or attributes. After a small number of transactions, the identity is known, even if it is never linked to exogenous information beyond the transactions themselves.

Previously, Facebook already undertook some endeavours to exchange data with banks. The Wall Street Journal observed how Facebook was approaching large US banks to share detailed financial information in exchange for "improving people's commerce experience". Seemingly, Facebook's interest was driven by its attempt to deepen user engagement and as a result, entice users to spend more time on Messenger [19]. Facebook executives felt the heat after announcing decreasing growth rates during their earnings call in 2018 that shaved more than $119 billion from its market value [20]. As a way to increase the use of its apps, and hence keep increasing its share price, Facebook will need to have contracts with partners that encourage access to their own services through Facebook's apps. If members of the Facebook ecosystem find a way to achieve this, they will be on their way to perfecting the fully-addressable 'transparent citizen'. Libra is a natural next step for Facebook, considering that

transaction data represents an untapped trove of high-quality personal data suitable for surveillance and control, as predicted already by Paul Armer in 1967 [21]. The business need for profit means that Facebook has few alternatives but to enter this market.

The 'transparent citizen' opens up further possibilities. Already in 2010, Facebook had bought a patent from now-defunct social networking site Friendster that would give creditors access to the social profiles of its users in order to assess them for a loan [22]. There is no evidence that it has yet been implemented, which could also be because of the outcry it received for seeking to control the friends and social behaviour of its users in order to provide a loan. Nevertheless, using social media information to provide more transparency about individual persons has already started to spread, and services that leverage Facebook data is in high demand. Yasaman Hadjibashi, chief data officer at Barclays Africa, explained in her interview with Business Insider: "*some of the world's biggest banks are working on using people's personal information from Facebook, Instagram, and other social media platforms as part of, or in lieu of, someone's credit history.*" She expresses her "*hope that everyone will one day have a unique code that will identify them and let people pull up all their information at once. However, as you can imagine, building the tech and infrastructure to enable this across jurisdictions and sectors is a mammoth feat*". [23]

This opinion is also reflected within the Libra association, as articulated by Shreves, senior adviser of technology at Mercy Corps, a founding member of the Libra association. He imagines that "a user's digital identity — for example, having an account with WhatsApp or Facebook — could be enough to be eligible." [24]. So, it seems some participants have been waiting for Facebook to fully enter the market as an authentication provider at the level of accuracy and trustworthiness that is required for financial institutions. Regardless of who will ultimately provide the financial service (for example it might actually be Mastercard, Visa or PayPal who are already members of the association), Facebook has positioned itself as the provider of the app that will provide the authentication of any user going forward and with all identities passing through their system.

## Digital Identity – a worthy business

Of course, the business of digital identity has not gone undiscovered. There are various attempts to estimate the market size for digital identification, authentication, and personal credentials. According to Juniper Research, an UK based analyst group, the number of people using government-issued Digital Identity is set to boom, rising by over 150% from an expected USD 1.7 billion this year to over USD 5 billion in 2024, with mobile single sign-up alone set to have over a billion users by 2023, and generate over USD 5 billion in revenues that year alone [25]. The Future of Personal ID to 2021 from Smithers Pira valued the global market for Personal ID credentials at USD 8.7 billion in 2016 and is forecast to reach USD 9.7 billion by 2021 [26]. Finally, Global Market Insights measured the market size around USD 6.5 billion in 2015 and is expected to reach about USD 34 billion by 2024 [27].

Whichever forecast is easier to believe, it seems apparent that there is already a strong market value with a strong growth estimate. Also, typically these analyses consider the use of digital identity for a person; however, the future will also include the need to provide trusted identity to devices as part of the big internet-of-things exchange of transactions.

Already, there have been multiple big digital players looking to solve this problem. Not surprisingly, they include some of the founding members of Libra, like MasterCard [28], Visa [29], and PayPal [30]. Microsoft is launching its decentralized identity infrastructure, called ION, built directly on the Bitcoin blockchain [31]. Also, Microsoft is doing this as part of DIF, an organization focused on developing the foundational elements necessary to establish an open ecosystem for decentralized identity and ensure interoperability between all participants. Currently DIF contains 79 members [32]. Christopher Allen, a crypto veteran and the co-founder of the World Wide Web Consortium (W3C) working group for decentralized identity (DID) solutions, told CoinDesk that Microsoft's move could impact the entire tech industry [33]. In addition, the crypto exchange unicorn Coinbase is exploring decentralized identity solutions [34], and there are a multitude of new players hoping to be the one big trusted partner or at least have a share of the cake [35]. Facebook has decided to not join these working groups and to pursue its own strategy through the Libra association.

Last but not least, Facebook executives are giving the answer. During the US House of Representatives Financial Services Committee, Markus was asked how Facebook will make money from Libra. Marcus said Facebook will benefit in two ways: First, the 90 million businesses on Facebook's platform will be able to make transactions with one another. Marcus predicted the increased commerce on the platform would help small businesses expand and ultimately spend more money on Facebook ads. But more importantly, second, "Facebook would offer more services in partnerships with banks and other organizations, from which it would expect to make money." This second business stream is where Facebook will become the authentication provider for many more businesses which we suggest would leverage its position as commercial digital identity provider [12].

## The winner takes it all

With all these big and small players having worked toward a solution for digital identity over multiple years, isn't it surprising how the Libra whitepaper throws it in as a side product not worth mentioning before Section 5? Facebook explained in June last year that 2.5 billion people used at least one of its apps: Facebook, Instagram, WhatsApp or Messenger [36]. This helpful number counts real people, although people can have multiple accounts on a single app. Two and a half billion people compares to 2.23 billion monthly users on Facebook, 1 billion users on Instagram, 1.5 billion users on WhatsApp and 1.3 billion users on Messenger [36]. Maybe Facebook mentions this endeavour so late because it will be so easy to win the game if you already speak directly to 2.5 billion users across multiple apps?

But Facebook's market power should not blend other institutions. Enabling a private advertising company to become the door-opener for most online services under standards and policies designed by them, can lead to disastrous outcomes as we have seen with Facebook before. Choosing a partner should be based on a thorough due-diligence that seeks to understand the motivation of its clients and includes reputational issues and wrong-doings that resulted in record-breaking fines [37]. Understanding future risks will also require data regulators to ensure awareness of risks that can further undermine the trust of citizens in the rule of law and its respective pillars, laws and bills to ensure rights for privacy, rights for freedom of opinion, rights for financial privacy, rights for fair credit reporting, and other local, regional and international rights. In addition, public safety administrators will need to investigate the economic and physical threats on the 'transparent citizen'. Given that these threats can also address whole nations, we must appeal also to national defence departments, given that we have already seen attacks from Russia during elections [38].

We submit that it would be wise not only to consider the requirements put forward to enable trusted digital financial services, but also to exercise particular vigilance when addressing the question of who a person is in the digital world, acknowledging that in the offline world, identity is a sovereign right traditionally only controlled by governments [39]. Handing this right over to a handful of selected private partners with a revenue-driven target could lead to biased decision-making and illegitimate gatekeepers for the sharing of information, a mechanism for using incentives, punishments, temptation, and fear to control the behaviours of populations, cheaply and at scale: a mix of Huxley's *Brave New World* and Orwell's *1984*. Following the argument offered by Alexandria Ocasio-Cortez: if money is a social good, then we should finally determine that the right of individuals to establish and maintain multiple, unlinked identities is a social good. And a social good should be managed by all participants from the private and public sector, including the individuals themselves. Blockchain can certainly provide good ideas, but an association of around 100 selected members, whose position compels them to obey the rules of one central private player with a shady history of securing peoples' data, is by no means getting the most and best out of it. In fact, it sounds like a charade.

## Acknowledgments

Valerie Khan is a co-founder and Vice-President of Digital Equity (http://d-eq.org), a Swiss based Association supporting all actors in the humanitarian and development space with their digital transformation in a responsible manner that seeks to serve everyone and harm no-one. She acknowledges her colleagues Dr. Richard Wilcox and Karl Steinacker for their thoughtful insights. Geoff Goodell is also an associate of the Centre for Technology and Global Affairs at the University of Oxford, and he acknowledges the Engineering and Physical Sciences Research Council (EPSRC) for the BARAC project (EP/P031730/1) and the European Commission for the FinTech project (H2020-ICT-2018-2 825215).